\documentclass[aps,prb,onecolumn,showpacs,showkeys,floatfix,groupedaddress]{revtex4-2}

\usepackage{amsfonts,fancybox,pgf}
\usepackage{amssymb}
\usepackage{float}
\usepackage{footnote}
	\usepackage{amsmath}
	\usepackage{makeidx}
	\usepackage{amsfonts}
	\usepackage[ansinew]{inputenc}
	\usepackage[usenames,dvipsnames]{pstricks}
	\usepackage{subfigure}
	\usepackage{graphicx}
	\usepackage{here}
    \usepackage{epstopdf}
	\usepackage{epsfig}
	\usepackage{pst-grad} 
	\usepackage{pst-plot} 
	\usepackage[colorlinks,hyperindex]{hyperref}
	\usepackage[english]{babel}

	\hypersetup
		{
		colorlinks,%
		citecolor=blue,%
		linkcolor=black,%
		urlcolor=black,%
	}

\makeindex

\begin{document}
\title{Generation of renormalized quadratic coefficient in Landau theory: Implications for specific-heat jump calculations in high-temperature superconductors}

\author{O.C. Feulefack$^{a}$, C. Tsague Fotio$^{b}$, R.M. Keumo Tsiaze$^{a,c ^{\dag}}$, S.E. Mkam Tchouobiap$^{a,d \S}$, J. E. Danga$^{b}$, A.J. Fotue$^c$, M.N. Hounkonnou$^{a}$}


\email{Corresponding Authors: keumoroger@gmail.com, esmkam@yahoo.com}

\affiliation{$^{a}$International Chair in Mathematical Physics and Applications, University of Abomey-Calavi, 072 B.P. 50, Cotonou, Republic of Benin.\\
$^{b}$Condensed Matter and Nanomaterials, Department of Physics, Faculty of Science, University of Dschang,P.O. Box 67, Dschang, Cameroon.\\
$^{c}$Laboratory of Mechanics, Materials and Structures, Department of Physics, Faculty of Science, University of Yaound\'e I - P.O. Box 812, Yaounde, Cameroon.\\
$^{d}$Laboratory of Research on Advanced Materials and Nonlinear Science (LaRAMaNS), Department of Physics, Faculty of Sciences, University of Buea, PO Box 63, Buea, Cameroon.\\}

\date{\today}

\begin{abstract}
In this work, Landau's theory is revisited by renormalizing quadratic coefficients derived from nonlinear polynomial equations to account for system dimensionality. In this respect, the generated coefficients, which include an intrinsic energy parameter specific to each material, enable precise specific-heat calculations for a range of high-temperature superconductors near the superconducting transition. To that end, the change in the specific heat jump is explained phenomenologically, which applies to any spatial arrangement and electron interactions that influence system symmetries. 
Moreover, effects leading to rapid, non-monotonic variation in the specific heat jump, $\Delta{C_p}/T_{c}$, across the transition are examined, with particular emphasis on changes attributed to the Sommerfeld coefficient in the normal state. The considerable reduction, disappearance, or significant enhancement of the specific heat anomaly at the superconducting transition is quantitatively explained by incorporating strong fluctuation corrections to the Landau theory for low-dimensional systems. 
Furthermore, the evolution of specific-heat jumps with system dimensionality is analyzed, and the results are discussed in relation to experimental observations of specific-heat jumps in yttrium- and bismuth-based superconductors, as well as in zero-dimensional superconductors.
\end{abstract}



\maketitle

\section{Introduction}
\noindent

Phase transitions and symmetry breaking in Landau-type models, first proposed by Landau and Ginzburg prior to the development of the BCS theory, have been extensively examined in the literature. Despite continued interest in this area, identifying a novel model that exhibits these properties remains challenging, given the extensive exploration already undertaken in this field. The BCS theory describes conventional superconductivity at low temperatures by explaining how phonons facilitate the formation of Cooper pairs. However, this theory does not fully account for high-temperature superconductivity. High-temperature materials exhibit characteristics such as $d$-wave pairing symmetry and strong electron-electron interactions, which extend beyond the simple $s$-wave pairing and phonon-mediated effects described by the BCS model \cite{Balatsky,Junod,Wesche,Bok}. In contrast, the Ginzburg-Landau theory (GLT) addresses scenarios in which the order parameter (OP) varies spatially, and the free energy depends on these variations. While BCS and GLT theories can account for the existence of a superconducting heat jump (SHJ), they do not explain its disappearance or the random variations arising from specific intrinsic material parameters. Several phenomena are associated with significant changes in the SHJ at the mean-field critical temperature $T_{c_0}$, including multiband superconductivity \cite{Penev, Kogan}, the presence of a normal state pseudogap \cite{JLoram}, and the effects of paramagnetic impurities \cite{Zaanen, Skalski}. Although significant theoretical work has been devoted to clarifying the pairing mechanism at $T_{c_0}$, the temperature dependence of the specific heat at this point remains unresolved. Currently, no universal theory explains the random behavior of the SHJ at $T_{c_0}$, which is influenced by the diverse configurations of materials. Nevertheless, several theoretical hypotheses have been proposed to account for certain experimental observations. 

Among  other things, calculations based on the GLT adequately describe the behavior of the electronic contribution $C_e/T_{c_0} = \gamma_0$ at $T_{c_0}$, where $\gamma_0$ represents the mean-field Sommerfeld coefficient of the normal state \cite{Gorkov}. However, in multiband superconductors, when both inter-band and intra-band pairing interactions are considered, significant deviations from the standard single-band GLT result are observed. These findings demonstrate that achieving a comprehensive understanding of the specific heat anomaly at the superconducting transition remains a complex challenge. High-$T_{c_0}$ superconductivity is likely governed by distinct pairing mechanisms, which may be associated with magnetic or other electronic interactions \cite{Junod,Wesche,Bok}. Also, the observed deviations from asymptotic behavior in several high-$T_{c_0}$ superconductors suggest that the GLT must be revised to more accurately account for newly observed phenomena at $T_{c_0}$ \cite{Wesche, Bok, Ferrell, Fisher, Gold}.

Reducing dimensionality leads to a loss of long-range order due to thermal fluctuations, as articulated by the Mermin-Wagner-Hohenberg theorem \cite{Mermin, Hohenberg}. While thermal fluctuations are suppressed over long distances in three-dimensional systems, they are more likely to generate topological defects, such as vortices, in two-dimensional systems \cite{Ktitorov, Herok}. Scalapino \textit{et al.} \cite{Ferrell} demonstrated that inadequate treatment of fluctuations underlies the failure of the GLT to account for microscopic interactions in physical systems. Although the GLT predicts a finite critical temperature for phase transitions, it does not determine the precise temperature at which the transition occurs; rather, it characterizes the system's behavior near this transition. In contrast, the Mermin-Wagner-Hohenberg theorem asserts that in condensed matter systems with short-range interactions and low dimensionality ($d \leq 2$), continuous symmetry cannot be spontaneously broken \cite{Mermin, Hohenberg}. Consequently, long-range order that would disrupt a continuous symmetry cannot exist in one- or two-dimensional systems. These findings suggest that the limitations of the GLT in explaining experimental results for low-dimensional materials are not solely related to the accuracy of critical exponents, as addressed by renormalization group theory \cite{Widom, Kleinert, RB, Fisher, Fisher2, Justin, Gold, Binney}. They also pertain to the interpretation of the transition temperature $T_{c_0}$, which should be regarded not as a true transition temperature but rather as a characteristic temperature scale associated with thermal fluctuations \cite{roger, roger2}. Therefore, any physical model of phase transitions must incorporate these principles to accurately represent material behavior as dimensionality decreases.

Within the GLT, the quadratic coefficient encapsulates essential physical information about the system. Rigorous characterization of this thermodynamic parameter enables the extraction of phenomenological insights, even without a detailed microscopic framework \cite{Zaanen, HohenbergKrekhov}. In the conventional GLT, the quadratic coefficient is typically associated with a single variable, specifically temperature. Previous research has shown that this coefficient may also depend on the system's dimensionality  \cite{roger3}. In contrast to conventional GLT, this study introduces a correction term for the quadratic coefficient and derives exact expressions that incorporate strong fluctuation effects, scattering-state contributions, and the influence of system dimensionality. The temperature- and dimensional-dependencies of the quadratic coefficient are characterized using nonlinear polynomial equations. These equations also account for anisotropy and intrinsic system characteristics by including a scale energy parameter, which quantifies the free energy barrier separating a metastable state from a state of lower free energy \cite{HohenbergKrekhov, roger4}. The quadratic coefficients in these nonlinear equations are adjusted to reflect both the system's dimensionality and the presence of significant fluctuations. Their behavior varies with dimensionality, spanning from zero-dimensional to high-dimensional cases. The renormalized coefficients are then applied to investigate both the disappearance and anomalous enhancement of the SHJ at $T_{c_0}$, incorporating strong fluctuation corrections to the OP within GLT. Strong OP fluctuations can induce strong-coupling superconductivity, which may be characterized by either a relatively high or a relatively low transition temperature, depending on the spatial arrangements and electron interactions that determine the system's symmetries. The SHJ may either vanish or increase in magnitude, depending on the crossover between different types of OPs that result from the interplay between thermal fluctuations and mass renormalization. This crossover, which occurs during the disappearance of the SHJ, results in a significant loss of entropy at $T_{c_0}$ \cite{WLoram, Skocpol}.

The structure of this paper is as follows. Section II introduces the GLT, discusses its limitations, and highlights the necessity for corrections by emphasizing the significance of dimensionality and intrinsic system parameters in transition analysis. Section III establishes a foundation for understanding the random behavior of the SHJ, particularly in relation to strong fluctuation corrections to the GLT, and supports the development of a dimensional phenomenology of critical behavior. This study serves as a resource for deriving alternative exact expressions for the specific heat that yield enhanced insights. The model's results improve simulation accuracy for the SHJ and enable precise prediction of specific physical constraints. Section IV compares the theoretical calculations with experimental data for yttrium- and bismuth-based superconductors, as well as in zero-dimensional superconductors. Finally, Section V presents the conclusion.

\section{Model Development and Methodology}
\noindent

As well known, from the thermodynamics standpoint Landau's theory provides a unified mean-field framework for describing phase transitions. When the symmetry group of the ordered phase is a subgroup of that of the disordered (most symmetric) phase, and long-wavelength fluctuations associated with the gradient term are neglected, the Landau free energy can be expressed as follows \cite{HohenbergPC, HohenbergKrekhov}:
\begin{equation}
H_{L}[|\phi|] = V\Big(\frac{r}{2}|\phi|^{2} + \frac{b}{4}|\phi|^{4}  - h|\phi|\Big),
\end{equation}
where $\phi$ denotes the OP, and which serves as the wave function for superconducting electrons. This parameter is expressed as a complex function $\phi(\textbf{r})$, comprising both real and imaginary components. Here, the variable $V$ indicates the system volume, while $h$ stands for the external applied field. In parallel, the coefficient $b > 0$ exhibits weak dependence on external thermodynamic parameters, including temperature and pressure. Importantly, the phase transition occurs when the quadratic coefficient $r$ changes sign. In this respect, the critical point can thus be estimated by setting the quadratic coefficient to zero, i.e., 
\begin{equation}
r = r_0(T - T_{c_0}) = 0.
\end{equation}
The relative temperature distance from $T_{c_0}$ is defined as $\tau = \frac{T - T_{c_0}}{T_{c_0}}$, which contrasts with the analogous quantity evaluated at the true critical temperature $T_c$. The primary limitation of Landau theory arises from its neglect of spatial fluctuations, which results from assuming a spatially uniform OP. The refinement of Landau's theory enables the development of a renormalization process for the quadratic and quartic coefficients. This process does not alter the structural conclusions of previous results but prompts a reassessment of mean-field approaches. To incorporate spatial fluctuations, the OP should be modeled as a tensor field rather than a constant. 

During a second-order phase transition, as the material is cooled, the OP increases continuously from zero, beginning at the true critical temperature. Employing a functional integral over a locally and continuously represented OP $\phi(\textbf{r})$ allows for the description of the partition function of materials near $T_{c_0}$. The extended nature of critical fluctuations in these conditions justifies interpreting $\phi(\textbf{r})$ as an average of localized magnetic moments in a lattice model over several lattice spacings \cite{Kleinert}. Consequently, the Ginzburg-Landau free energy functional for high-$T_c$ superconductivity in the absence of an external field can be formulated as follows \cite{HohenbergKrekhov, roger, Ausloos, roger5}:
\begin{equation}
H_{GL}[|\phi|] = \int d^{d}\textbf{r}\bigg[\frac{\mathcal{C}}{2}\big(\nabla_r|\phi|(\textbf{r})\big)^2 + \frac{r(T)}{2}|\phi|^{2} + \frac{b}{4}|\phi|^{4}  +  \cdots \bigg].
\end{equation}
The coefficient $\mathcal{C}$, analogous to $b$, exhibits weak dependence on temperature and pressure and can be determined from microscopic theory for a given system. Characterization of any physical system is possible by selecting suitable coefficients in Landau's expansion, provided that a sufficient number of terms and relevant physical factors are considered. However, precise evaluation of specific thermodynamic parameters within the Ginzburg-Landau model can be challenging. Modeling interactions using the Hartree-Fock approximation \cite{roger5, Masker, Tucker} results in renormalization of the quadratic coefficient by terms reflecting the system's dimensionality, correlation functions, and scattering states. While the standard formulation assumes a linear temperature dependence of the quadratic coefficient near the transition temperature, rigorous mathematical derivations introduce corrections that make its behavior depend on the system's specific properties. The present analysis demonstrates the dual dependence of the quadratic coefficient on both temperature and dimensionality, as characterized by nonlinear polynomial equations. The resulting intrinsic parameters are then interpreted in terms of their physical significance.

\subsection{Expressions of renormalized quadratic coefficients}
\noindent

To describe a finite-temperature phase transition, a set of weakly interacting particles or quasi-particles is considered. The $\phi^2$ term is associated with essential fluctuations, whereas the $\phi^4$ term is linked to redundant fluctuations; thus, the quartic term serves as an interaction among the Fourier components of the OP. Since low-dimensional systems cannot establish long-range order at finite temperatures due to fluctuations in the OP, the approximation $|\phi(q)|^4 \approx 6\langle|\phi(q)|^2\rangle|\phi(q)|^2$ assumes that the Fourier components interact only through the mean field produced by other modes \cite{roger2, roger5}. Employing a cutoff value $\Lambda$ corresponding to lattice periodicity, fluctuations with wavelengths greater than the sample's coherence length are considered negligible. This decoupling is justified when the quartic coefficient in the expansion is small \cite{roger}. By coherently linking fluctuation strength and dimensionality, this model provides a promising framework for interpreting complex superconducting phenomena, resulting in a non-linear temperature dependence with quadratic coefficients as described by the following  transformation \cite{roger2, roger5}:
\begin{equation}
r_0(T - T_{c_0}) \rightarrow r_0(T - T_{c_0}) + \Sigma(d, T) \phantom{..}, \textrm{where} \phantom{..} \Sigma(d, T) = \eta\left(\frac{T}{T_{c_0}}\right)\left(\tau +\frac{\Sigma(d, T)}{r_{0}T_{c_0}}\right)^{\frac{d}{2}-1}.
\end{equation}

Here, the variational energetic parameter associated with the harmonic variance of the OP is represented by the correction quantity $\eta$ and which accounts for the strong dependence of the temperature and magnitude of the fluctuation corrections on the anisotropy of the electron spectrum and the potentially associated anomalies displayed by superconducting properties. Its amplitude is dependent on the coherence length as well as the fourth order coupling constant, which is known to influence the saturation of ordering in the GLT and account for the system's self-interaction \cite{roger,Ausloos,Larkin,Varlamov}.
More importantly, it should be noted that the dimensional-dependent quantity $\Sigma(d, T)$, as derived in Ref. \cite{roger}, is self-consistent and of great importance for the renormalization of the quadratic coefficient in Landau theory. Incorporating these corrections allows the Ginzburg-Landau theory to more accurately characterize phase transitions, thereby generalizing GLT to account for a wide range of thermodynamic anomalies via a unified fluctuation-driven mechanism. The form of $\Sigma(d, T)$ enables the reformulation of the new quadratic coefficients as a non-linear polynomial equation \cite{roger3, roger4} in the following form:
\begin{equation}
r(d, T) - \eta\left(\frac{T}{T_{c_0}}\right)\Bigg[\frac{r(d, T)}{r_0T_{c_0}}\Bigg]^{d/2 - 1} - r_0(T - T_{c_0}) = 0.
\end{equation}
A noteworthy matter is that previous studies generally utilized numerical methods to model thermodynamic functions. In contrast, the present analysis focuses on obtaining the exact solution of Eq. (5) to determine the renormalized quadratic coefficients. These coefficients facilitate the calculation of specific heat expressions as functions of dimensionality. Analysis of Eq. (5) indicates that fluctuations increase with $r(d, T)$ raised to the power $d/2 - 1$, provided that fluctuation corrections are not negligible for $d \leq 2$. The dimension $d_c = 2$ represents a critical threshold below which fluctuations become highly significant. It is necessary to specify a range of possible values for $\eta$ in various representative experimental scenarios. Solving the nonlinear polynomial Eq. (5), which involves the variables dimension $d$ and temperature $T$, typically requires substitution or elimination to reduce the equation to a single variable. Assigning a specific value to the system's dimension allows for analysis of the temperature-dependent behavior of thermodynamic functions. Solutions to Eq. (5) for each dimension yield the following quadratic coefficients:
\begin{eqnarray}
&&r(d, T) =  \left\{ \begin{array}{lllll}\vspace{0,5cm}
\frac{1}{2}\Big(r_0(T - T_{c_0}) \pm \sqrt{r^2_0(T - T_{c_0})^2 + 4\eta{r_0}T}\Big), & \textrm{for}\phantom{.} d = 0  \\ \vspace{0,5cm}
r(d = 1, T), & \textrm{for}\phantom{.} d = 1  \\ \vspace{0,5cm}
\eta\left(\frac{T}{T_{c_0}}\right) + r_0(T - T_{c_0}), & \textrm{for} \phantom{.} d = 2\\\vspace{0,5cm}
\frac{\eta{T}\Big(\eta{T} \pm\sqrt{\eta^2T^2 + 4r^2_0T^3_{c_0}(T - T_{c_0})}\Big)}{2r_0T^3_{c_0}} + r_0(T - T_{c_0}), & \textrm{for} \phantom{.} d = 3\\ \vspace{0,5cm}
r_0\big(1 - \frac{\eta{T}}{r_0T^2_{c_0}}\big)^{-1}(T - T_{c_0}), & \textrm{for} \phantom{.} d = 4\\
  \end{array}\right.
\end{eqnarray}
Consequently, Eq. (6) demonstrates that a dimensional Ginzburg-Landau free energy functional can be derived by integrating out the OP fluctuations exactly and by treating interactions within the system using a self-consistent approximation \cite{Ausloos}. The function $r(d = 1, T)$ remains positive for all temperatures. Although the explicit expression is complex and not presented here, its behavior is depicted in Fig. 1 for $\eta$ = 0.85. The value $r_{1D}(T)$ becomes zero at $T$ = 0 K for any value of $\eta$. Analysis of Equation (6) indicates that critical parameters and exponents are functions of the system's dimensionality and display quantum characteristics in low-dimensional systems.
\begin{figure}
 	\centerline{\includegraphics[width=10cm]{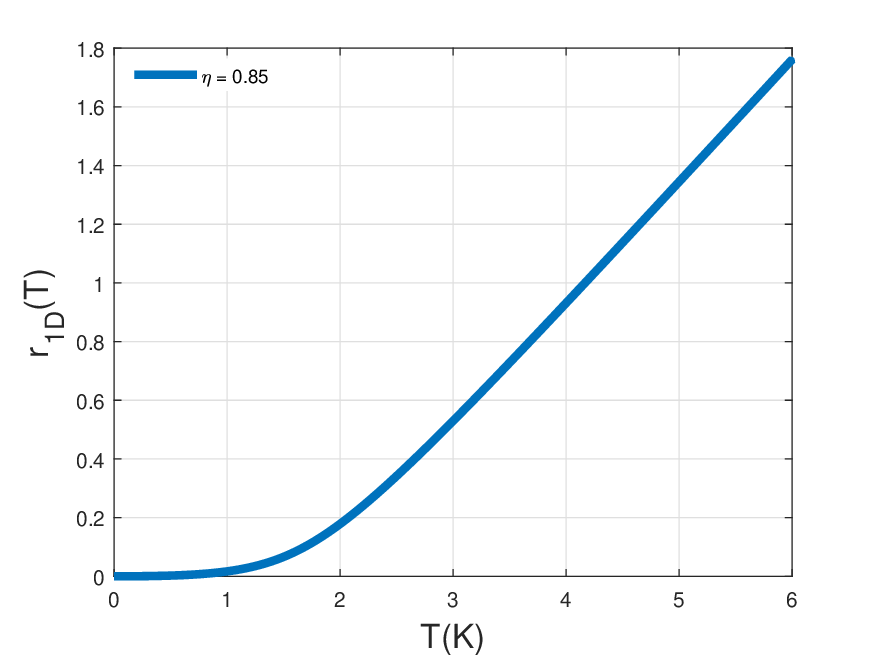}}
 	\caption{Temperature dependence of the renormalized one-dimensional quadratic coefficient as described by Eq. (5). The coefficient $r_{1D}(T)$ remains positive at all temperatures and becomes zero at $T$ = 0 K for any value of $\eta$.}
 	\label{figure1}
 \end{figure}
Mathematical analysis of Eq. (6) indicates that increasing the magnitude of $\eta$ suppresses phase transitions in the system for $d = 0$ and $d = 1$. While the new quadratic coefficients continue to cancel at a finite transition temperature, the associated critical behavior is modified in both qualitative and quantitative terms. Accordingly, the modified critical temperature can be expressed as follows:

\begin{eqnarray}
&&r(d, T) = 0 \Rightarrow  \left\{ \begin{array}{lllll}\vspace{0,7cm}
T_{c_{0D}} = 0, & \textrm{for}\phantom{.} d = 0  \\ \vspace{0,7cm}
T_{c_{1D}} = 0, & \textrm{for}\phantom{.} d = 1  \\ \vspace{0,7cm}
T_{c_{2D}} = T_{c_0}(1 + \frac{\eta}{r_0T_{c_0}})^{-1}, & \textrm{for} \phantom{.} d = 2 \\ \vspace{0,7cm}
T_{c_{3D}} = T_{c_0}, & \textrm{for} \phantom{.} d = 3\\ \vspace{0,7cm}
T_{c_{4D}} = T_{c_0}. & \textrm{for} \phantom{.} d = 4\\
  \end{array}\right.
\end{eqnarray}
It is noteworthy that Eq. (7) integrates both dimensionality and thermal fluctuations into the theoretical framework for critical temperature, thereby demonstrating that size effects significantly affect the thermodynamic properties of materials. The critical temperature, as a function of system dimensionality, displays non-monotonic behavior. The temperature at which $r(d = 2, T)$ changes sign decreases as $\eta$ increases and is strongly determined by the nature of fluctuations within the system. For sufficiently large $\eta$, $T_{c_{2D}}$ approaches zero, in agreement with the Mermin-Wagner theorem. The model also indicates that, due to fluctuations in the OP, one-dimensional systems and two-dimensional systems with higher $\eta$ cannot establish long-range order at finite temperatures. Additionally, the coefficients $r_0$ and $\eta$, as derived from microscopic theories, suggest that fluctuations remain significant over a temperature range comparable to the mean-field transition temperature for dimensions $d \leq 2$ \cite{Mckenzie, Pettini}.

\subsection{Physical interpretation of the energy parameter $\eta$}
\noindent
As presented above and established known from previous studies, the parameter $\eta$ serves as an intrinsic coefficient within the model, representing a fundamental property unique to a specific material \cite{roger,roger2,roger5}. It defines an intrinsic energy parameter associated with resistance between Cooper pairs, spins, or quasi-particles, depending on the system under consideration. The energy parameter $\eta$ is typically defined as a scalar quantity with the same dimension as the quadratic coefficient $r_0(T-T_c)$. It plays a critical role in preventing transitions from high- to low-temperature phases at finite temperature, as increasing fluctuation corrections gradually result in the annihilation of the low-temperature order. Equation (7) describes the temperature dependence of the true critical temperature as a function of the dimension $d$. Within the two-dimensional framework, the scaled fluctuation-correction term $\varepsilon$ is introduced as $\varepsilon = \frac{\eta}{r_0T_{c_0}}$, which defines the relative temperature with respect to the mean-field critical temperature $T_{c_0}$. In two-dimensional systems, significant critical fluctuations are recovered, and the true critical point $T_{c_{2D}}$ varies between 0 K and $T_{c_0}$, such that $0 \leq T_{c_{2D}} \leq T_{c_0}$. The origin of superconductivity in two-dimensional materials remains a subject of debate. Some arguments are based on the Fermi-liquid framework, while others propose that superconductivity may arise from a non-Fermi-liquid quantum-critical metal \cite{SKim}. Upon cooling a superconducting material, certain electrons participate in the pairing mechanism, whereas others contribute to charge transfer as one-electron excitations. Single- or one-electron excitations are generally associated with spin-density wave order \cite{Doren}, which is essential for understanding the magnetic and electronic properties of high-$T_{c_0}$ cuprates, particularly those exhibiting density wave phenomena. A spin density wave OP characterizes the spatial modulation of electron spins, resulting in a periodic pattern of spin density \cite{Vavilov, Sil}. In this study, $\eta \propto g(E_{\eta})$ is related to the density of states for single-electron excitations, specifically the density of states of unpaired electrons. In contrast, $r_0T_{c_0}$ denotes the density of states of electrons that contribute to superconductivity. In contrast to the classical electronic density of states, $g(E_{\eta})$ represents the spatial distribution or energy-level availability of unpaired spins within a system. Localized, unpaired electrons, which are prevalent in overdoped superconductors, radicals, and transition metals, can be described using electron density matrices. This approach enables the analysis of antiferromagnetic coupling and radical character \cite{Lain}. Experimental evidence supports the coexistence of superconducting and non-superconducting (normal) regions, as a subset of electrons does not form Cooper pairs. In overdoped $La_{2-x}Sr_xCuO_4$, for example, the presence of unpaired electrons in the overdoped regime is a significant indicator that the system is transitioning from a strong-pairing superconducting state to a conventional metallic state. The superfluid density decreases in a manner not fully accounted for by standard BCS theory \cite{Hijano, Long}.

In parallel, the pairing mechanism facilitates the formation of Cooper pairs via attractive interactions, leading to positive coupling. In contrast, one-electron excitations induce repulsive interactions, which counteract the pairing mechanism. These excitations delay the onset of pairing, leading to antagonism between the two processes and this antagonism reflects the coexistence and interplay of competing phases within the material. A noteworthy matter is that each phase is characterized by a spontaneously broken symmetry, leading to strong fluctuations and unsynchronized electron behavior that fundamentally alter the system's physics. Henceforth, the thermal evolution of different OPs demonstrates coexistence and re-entrance behaviors consistent with experimental observations \cite{Sil, Kitagawa}. By preventing a transition at a finite critical temperature, in contrast to the pairing mechanism at $T_{c_0}$, one-electron-density wave order shifts the transition point and promotes the formation of topological defects such as vortices \cite{Ktitorov}. In 0D- and 1D systems, one-electron excitations largely dominate over the pairing mechanism; however, the absence of confinement or increased dimensionality favors the pairing mechanism. More details will be discussed later. The specific heat versus temperature curves are predicted to be nonuniversal and system dependent, which precludes direct comparison of the present results to a quantitative model. Additionally, the relationship between the energy parameter $\eta$ and spin-density wave order remains insufficiently characterized, underscoring the need for further comprehensive investigation

 \section{Renormalized specific-heat jump calculations}
\noindent

This section derives a priori and a posteriori fluctuation-correction estimates based on the exact calculation of quadratic coefficients within Landau theory. The analysis emphasizes the dual dependence of these estimates on both dimension and temperature, and their implications for specific-heat jump calculations.

As well etablished, the GLT provides a foundational mathematical and physical framework for modeling phase transitions, particularly in superconductivity \cite{roger2, Ktitorov}. Its applications encompass both conceptual modeling and computational analysis to achieve accurate approximations of solutions. Following the decoupling of the quartic term to renormalize the standard quadratic coefficient, a comparable approach can be applied to the sixth-order term through approximations or specific mathematical transformations in order to restore the quartic term. This approach facilitates the analysis of complex equations, supports the development of more precise theories in quantum mechanics, and simplifies highly nonlinear systems by rendering their components independent. The renormalization process outlined here provides a systematic method for examining the evolution of fundamental parameters with scale or energy. Incorporating nonlinear corrections into conventional Landau theory enables the characterization of long-distance phenomena while accounting for short-distance physics. Using the transformation of Eq. (4), the effective Landau free energy takes the following form \cite{HohenbergPC}:
\begin{equation}
H_{eff}[|\phi|] = V\Big(\frac{r(d, T)}{2}|\phi|^{2} + \frac{b^{\star}}{4}|\phi|^{4}  - h|\phi|\Big).
\end{equation}
The equilibrium value of the renormalized OP is established by minimizing the Landau functional and the renormalized specific heat $C_{p} = -\frac{T}{V}\frac{\partial^2{H_{eff_{Min}}}}{\partial{T^2}}$ in the immediate neighborhood of the true critical temperature $T_{c}$ can be factorized as follows:
\begin{eqnarray}
&&C_{p}  =  \left\{  \begin{array}{ll}\vspace{0,5cm}
C_0(T), & \textrm{for}\phantom{...} T > T_{c}, \\
C_0(T) + \Delta{C_p}(d, T), & \textrm{for} \phantom{...} T < T_{c}.\\
  \end{array}\right.
\end{eqnarray}
Here $C_0(T)$ represents a function that characterizes the normal state electronic specific heat, while $\Delta{C_p}(d, T)$ defined as
\begin{equation}
\Delta{C_p}(d, T) = \Bigg[\bigg(1 + T_{c_0}\frac{\partial{\upsilon(d, T)}}{\partial{T}}\bigg)^2 + T^2_{c_0}\Big(\tau + \upsilon(d, T)\Big)\frac{\partial^2{\upsilon(d, T)}}{{\partial{T}}^2}\Bigg]\frac{r^2_0T}{4b},
\end{equation}
denotes the abrupt condensation of electrons into paired states, thereby limiting the degrees of freedom accessible for thermal excitation. 

More importantly, the dimension- and temperature-dependent quantity $\upsilon(d, T)$
\begin{equation}
\upsilon(d, T) = \frac{\Sigma(d, T)}{r_0T_{c_0}} =  \frac{r(d, T)}{r_0T_{c_0}} - \tau
\end{equation}
establishes the sequence of fluctuations in the system, facilitating the transition from classical mean-field critical behavior [$\upsilon(d, T) \rightarrow 0$] to fluctuation-dominated critical behavior [$\upsilon(d, T) \neq 0$].

\subsection{Weak-Fluctuation Regime: $\upsilon(d, T) \rightarrow 0$}
\noindent

The observed discontinuity arises because the Cooper pair condensate has lower entropy than the normal electron state, leading to an abrupt change in the temperature derivative of entropy at the mean-field (MF) critical temperature. In the mean-field regime ($\upsilon(d, T) \rightarrow 0$), the SHJ is given by
\begin{equation}
\Delta{C_{MF}}(T_{c_0}) = \frac{r^2_0T_{c_0}}{4b}.
\end{equation}
Just as with the BCS theory, the Landau model implies that the SHJ is proportional to $T_{c_0}$ through a scaling behavior $\Delta{C_p} = \gamma_0{T}_{c_0}$. Here the MF Sommerfeld coefficient of the normal state $\gamma_0$ given by 
\begin{equation}
\gamma_0 = \frac{r^2_0}{4b}.
\end{equation}
The Sommerfeld coefficient quantifies the electronic specific heat at low temperatures and directly reflects the density of states at the Fermi level.

\subsection{Strong-Fluctuation Regime: $\upsilon(d, T) \neq 0$}
\noindent

Thermal fluctuations can alter the sharp jump, often resulting in a rounded discontinuity. At the transition temperature in certain high-$T_{c}$ superconductors, the behavior of the SHJ varies, ranging from complete disappearance \cite{Tanaka, WLoram} to exponential growth \cite{roger}, as well as exhibiting scaling behavior described by $\Delta{C_p} = AT^n_{c}$, where $A$ is a constant and $n$ is the scaling exponent. For $n = 3$, this scaling reflects the thermodynamic properties of systems with strong critical quantum interactions, where the electrons responsible for superconductivity contribute a $T^3$ term to the specific heat. The gradual increase in $\Delta{C_p}$ at $T_{c}$ arises because the superconducting gap opens exponentially rapidly \cite{Zaanen, Prozorov}. In electron-doped compounds, particularly within the $Ba(Fe_{1 - x}Tm_x)_2As_2$ family with $Tm = Co, Ni$, the case of $n = 2$ corresponds to the ideal Bud'ko, Ni, and Canfield (BNC) scaling \cite{Bang}, which is limited to a narrow range of $T_{c}$ variation \cite{Grinenko, Stewart}. According to the theories of Landau and BCS, $\Delta{C_p} = \gamma_0T_{c}$, indicating a scaling exponent $n$ = 1.
\begin{figure}
 	\centerline{\includegraphics[width=8cm]{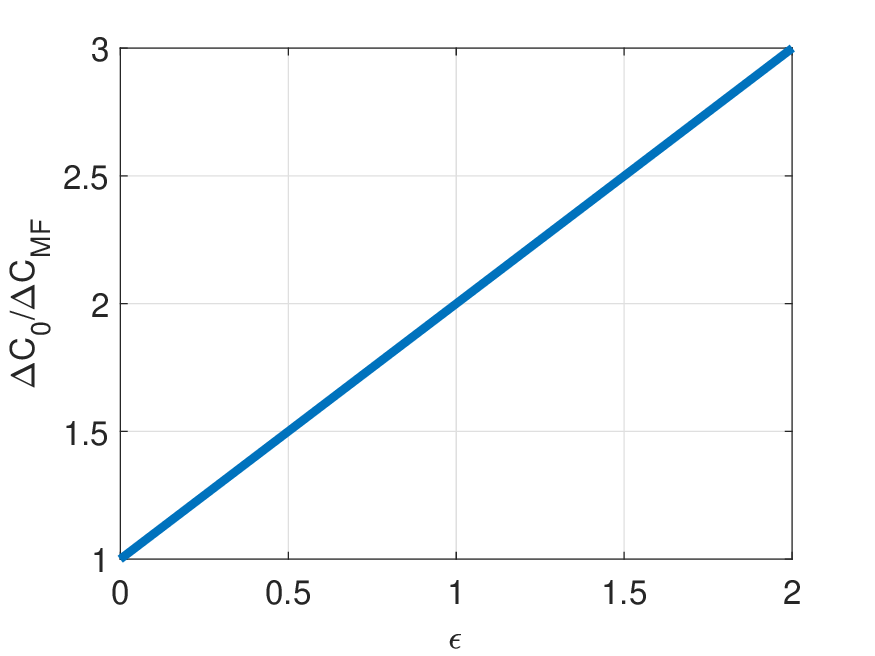}}
 	\caption{Plot of the ratio between the renormalized and standard specific-heat jumps as a function of the scaled fluctuation-correction term $\varepsilon$ for a two-dimensional system. The renormalized SHJ increases linearly in comparison to the standard SHJ as fluctuation effects increase.}
 	\label{figure1}
 \end{figure}
 
\begin{figure}
 	\centerline{\includegraphics[width=8cm]{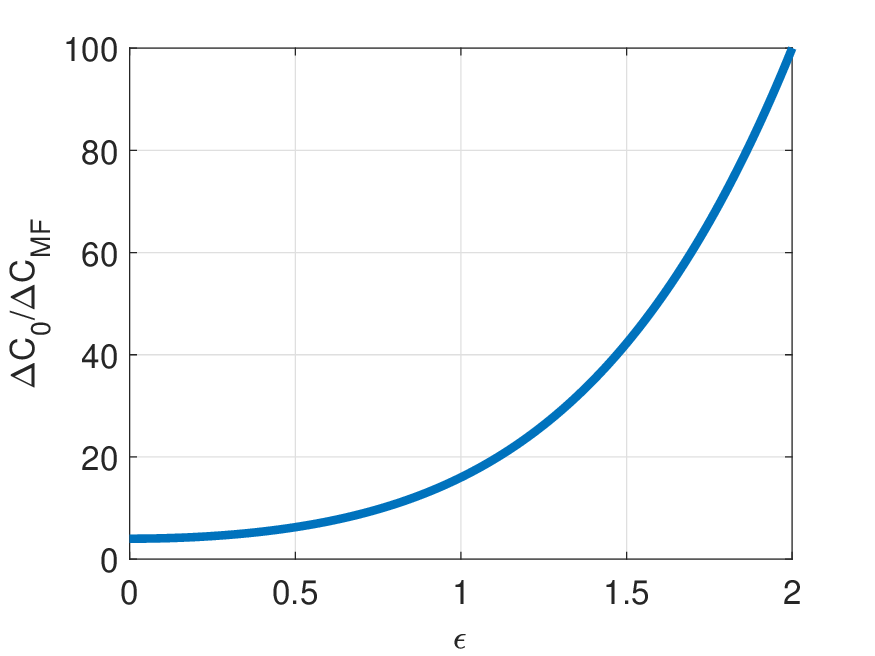}}
 	\caption{Plot of the ratio between the renormalized and standard specific-heat jumps as a function of the scaled fluctuation-correction term $\varepsilon$ for a three-dimensional system. The renormalized specific-heat jump increases quasi-exponentially in comparison to the standard specific-heat jump as fluctuation effects in the system increase.}
 	\label{figure2}
 \end{figure}
 
Eq. (10) is intended to characterize various forms of specific-heat jump behavior, particularly during the superconducting transition and, in some cases, during general phase transitions. This equation incorporates dimensional scaling theory as applied to strongly coupled or unconventional superconductors. Several practical cases can be demonstrated using the expressions for the renormalized quadratic coefficient provided in Eq. (6).

\begin{eqnarray}
 \frac{\Delta{C_0}}{\Delta{C_{MF}}} = \left\{  \begin{array}{ll}\vspace{0,4cm}
 
(1 + \varepsilon), & \textrm{for} \phantom{...} d = 2 \\\vspace{0,4cm}

0 \phantom{..} \textrm{or} \phantom{..} 4(1 + \varepsilon^2)^2, & \textrm{for} \phantom{...} d = 3. \\
  \end{array}\right.
\end{eqnarray}
As essential contribution, Eq. (14) defines the ratio of renormalized to standard specific heat jumps as a function of the scaled fluctuation correction term $\varepsilon$ for both two- and three-dimensional systems. The corresponding behaviors are depicted in Figs 2 and 3. Fig. 2 shows that, the renormalized specific-heat jump in 2D systems increases linearly relative to the standard value as fluctuation effects intensify. Eq. (5) demonstrates that, in 3D systems, two distinct solutions exist for the quadratic coefficient. The first solution produces a quasi-exponential increase in the SHJ as the scaled fluctuation-correction term $\varepsilon$ increases, as shown in Fig. 3. A small increase in $\varepsilon$ in this particular case leads to the divergence of specific heat at the critical point, rather than a discontinuous jump. In contrast, calculations based on the second solution yield no SHJ at $T_{c_0}$. This absence of a jump is physically attributed to the presence of a partial gap (pseudogap) in the electronic density of states at the Fermi level. This phenomenon, which is characteristic of underdoped cuprates, may result from several factors: (i) impurities that broaden the superconducting transition and influence the SHJ, leading to competition between pseudogap and impurity effects \cite{Dzhumanov}; (ii) strong electron-phonon coupling \cite{YWang}; or (iii) anisotropic superconductivity, particularly in underdoped cuprates, where the superconducting gap is anisotropic and produces a complex specific-heat response.

Accordingly, and of crucial importance, in Eq. (10) the factor $(\tau + \upsilon)$ becomes zero at the transition point. This condition enables the definition of the renormalized Sommerfeld coefficient of the normal state as follows:
\begin{equation}
\gamma(d, T_{c_{d}}) = \gamma_0\Bigg[\bigg(1 + T_{c_0}\frac{\partial{\upsilon(d, T)}}{\partial{T}}\Big\vert_{T = T_{c_d}}\bigg)^2 \Bigg].
\end{equation}
Strong interactions within the lattice increase the effective mass of charge carriers, thereby enhancing the Sommerfeld coefficient by a factor that depends on dimensionality. Eq. (15) indicates that this coefficient is strongly determined by the intrinsic material parameter and establishes a universal framework for analyzing rounding behavior, as well as the fluctuation-dependent linear and quasi-exponential divergence of the specific-heat jump, as presented in Figs. 2 and 3. A quasi-exponential form of the SHJ is observed in specific families of $\beta$-pyrochlore oxide superconductors, which display a Sommerfeld coefficient $\gamma$ of exceptionally large magnitude, greatly surpassing values typical of conventional superconductors \cite{Bruhwile, Yonezawa}. The parameter $\upsilon(d, T)$, derived from the Hartree-Fock approximation, governs the scattering mechanisms within the material.

The present model reflects a deliberate methodological shift from a power-law approach to a coefficient renormalization framework. Rather than characterizing the specific heat jump, $\Delta{C_p}/T_{c}$, across the transition using a power law to capture scale-invariant phenomena, the methodology emphasizes adjusting internal model coefficients according to the scale of observation. This approach indicates a structural, mechanistic model rather than a purely empirical or observational fit.

\section{Comparison to experiment}
\noindent 

Single-electron excitations, fluctuations, and the specific-heat jump are closely related phenomena in condensed-matter physics, especially in the study of superconducting transitions. These concepts characterize the mechanisms by which electrons absorb energy, the system's behavior near critical points, and the thermodynamic indicators of the onset of superconductivity. During the transition to a superconducting state, the rearrangement of electrons and the emergence of an energy gap significantly modify the density of states near the Fermi level. When a superconductor is cooled below the mean-field critical temperature $T_{c_0}$, some electrons participating in pairing undergo charge transfer as single-electron excitations. As the temperature approaches the true critical temperature, the number of electrons involved in superconductivity gradually decreases, in contrast to the abrupt drop at $T_{c_0}$ observed in standard Landau theory. Single-electron excitations serve as direct evidence of a spin-density wave order induced by thermal fluctuations \cite{Vavilov}. In unconventional superconductors, particularly near a superconductor-insulator transition or in high-temperature cuprates, resistance between preformed Cooper pairs, often attributed to phase fluctuations or pair scattering, can suppress the expected specific-heat jump at the critical temperature. This suppression is associated with preformed Cooper pairs that exist above the transition temperature but lack long-range coherence. This phase fluctuations of these pairs inhibit the sharp thermodynamic transition observed when global phase coherence is established. 
The specific-heat jump marks the opening of a gap to single-electron excitations, while fluctuations around that transition (due to pairing or quantum criticality) determine the shape, magnitude, and scaling of this anomaly, often changing the mean-field behavior into a more complex, strongly correlated, or lower-dimensional behavior. In the framework of our model, there is no SHJ in zero- and one-dimensional cases. Shapes in these systems are distinct and exhibit characteristic properties of materials with finite energy levels. There is also zero SHJ in some bulk systems under particular conditions. By incorporating all relevant information from the new quadratic coefficients, the following expressions have been derived to describe the variation of the Sommerfeld coefficient in the normal state at the true critical point for 2D and 3D systems:

\begin{eqnarray}
 \gamma(d, T_{c_{d}}) = \left\{  \begin{array}{ll}\vspace{0,4cm}
 
\gamma_0(1 + \varepsilon), & \textrm{for} \phantom{...} d = 2 \\\vspace{0,4cm}

0\phantom{..} \textrm{or} \phantom{..} 4\gamma_0(1 + \varepsilon^2)^2, & \textrm{for} \phantom{...} d = 3. \\
  \end{array}\right.
\end{eqnarray}
In this section and to validate the inference framework, we demonstrate the usefulness of the obtained results in comparison with experimental results observed for the specific jumps of some materials such as, (a) Yttrium-based superconductors, (b) Bismuth-based superconductors, as well as (c) zero-dimensional superconductors whose specific-heat jumps do not belong to the same universality class.
 
\subsection{Specific jump of Yttrium-based superconductors}
\noindent 

By comparing the results of our theoretical model with experimental data on YBCO, we assess the alignment between observed outcomes and our model's predictions. This comparison is essential for validating and verifying scientific claims. In YBCO synthesized as $YBa_2Cu_3O_{7-\delta}$ with $0 \leq \delta \leq 0.18$, variations in oxygen stoichiometry influence the density of states at the Fermi level and, consequently, the magnitude of the SHJ $\Delta{C_p}$ \cite{Cooper, DFisher}. The coherence length, or separation of Cooper pairs, is approximately $\xi_0 \simeq 1$ nm in YBCO, which is much smaller than the typical value of $\xi_0 \simeq 1000$ nm in conventional superconductors. This reduced coherence length indicates that Cooper pairs are more localized, likely due to changes in oxygen stoichiometry. The localization of Cooper pairs, combined with the pronounced difference between the out-of-plane $\xi_{\perp}$ and in-plane $\xi_{\parallel}$ coherence lengths, results in a broad temperature range where critical fluctuations are observed in YBCO \cite{Seki,Jung}. The ratio $\rho = \frac{\xi_{\parallel}}{\xi_{\perp}}$ is a key parameter in determining the temperature range of critical fluctuations and the characteristics of OP fluctuations. Depending on the value of $\rho$, the system may enter different critical regimes. YBCO can also transition from a 2D to a 3D state under external influences such as strong magnetic fields or uniaxial pressure, which promote the formation of a 3D charge density wave. In terms of growth patterns, YBCO films transition from 2D layer-by-layer growth to 3D growth after exceeding a critical thickness, a process influenced by lattice mismatch and accumulated strain energy \cite{Osquiguil}. Models such as the 3D-XY and 2D-Anderson-Lawrence (2D-AL) are employed separately to study YBCO, reflecting its anisotropic properties \cite{Bok,Junod}. The 3D-XY model describes three-dimensional superconducting behavior with strong interlayer coupling, while the 2D-AL model addresses behavior within a single uncoupled layer or weakly coupled layers. 

On the other hand, the proposed model demonstrates excellent characteristic properties as a general-purpose framework, with specific applications determined by the YBCO configuration\cite{Bok,Jung}. This approach enhances efficiency and scalability in managing multiple specialized models. The dimensionality of the OP is analyzed to characterize the crossover from 2D to 3D behavior, where 2D behavior serves as a precursor to 3D superconducting order. Comparing the specific heats of 3D (bulk) and 2D (thin-film or layered-structure) systems is essential for elucidating fluctuation phenomena and the superconducting phase transition. Figure 4 illustrates the crossover from 3D to 2D behavior, which is accompanied by a reduction in the specific-heat jump. These observations are qualitatively consistent with the 3D-XY and 2D-AL predictions in the limit of an ultrathin, uniform superconducting film. The increase in the ratio between the 2D and 3D specific-heat jumps for $0 \leq \varepsilon \leq 0.2$ reflects a complex process involving increased rate-domain lifetimes and a transition from a homogeneous to a heterogeneous regime. Specific heat measurements in 2D YBCO or highly anisotropic thin-film YBCO systems exhibit a characteristic anomalous jump at the superconducting transition temperature, often with significant and broad fluctuation contributions that distinguish these systems from 3D mean-field superconductors. The specific-heat jump, defined by a discontinuity or anomaly in $\Delta{Cp}/T_c$, manifests as a rapid rise upon cooling through $T_c$ and is typically characterized by strong positive curvature on both sides of the transition.

\begin{figure}
 	\centerline{\includegraphics[width=8cm]{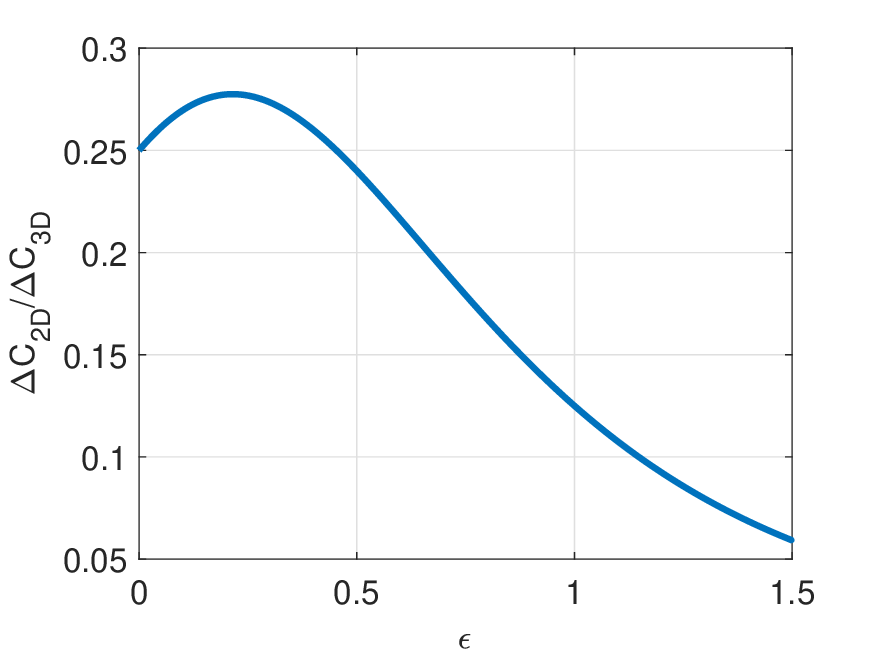}}
 	\caption{Plot of the ratio between the 2D and 3D specific-heat jumps as a function of the scaled fluctuation-correction term $\varepsilon$. The specific heat jump in the two-dimensional system may be up to eight times lower than that observed in the bulk material for low values of $\varepsilon < 1$. The methodology underlying the model requires the inclusion of a low-magnitude fluctuation-correction term}
 	\label{figure7}
 \end{figure}

\subsection{Specific jump of Bismuth-based superconductors}
\noindent 

A zero jump indicates the absence of a direct relationship among the specific heat jump, the density of states, and the transition temperature. This observation raises important questions regarding the nature of superconductivity and the role of the pseudogap, underscoring the structural complexity of cuprates and the necessity for more advanced models to account for their thermodynamic properties. Such behavior is characteristic of strong-coupling or unconventional superconductors. The principal advantage of this model, compared to existing theories for both conventional and unconventional superconductors, is its complete predictability. For example, with the ratio $\rho = \frac{\xi_{\parallel}}{\xi_{\perp}} \sim 30$, BiSrCaCuO exhibits a highly anisotropic crystal structure with layered CuO$_2$ planes that localize superconducting charge carriers. In contrast, YBCO$_{7 - \delta}$, where $\rho \sim 5$, demonstrates a relatively greater number of effective degrees of freedom than BSCCO. Consequently, quasi-two-dimensional effects become significant, and Cooper pairs decay more rapidly as temperature increases near the critical point. As discussed in subsection III, the zero-jump can be physically explained by the presence of a partial gap in the electronic density of states at the Fermi level \cite{JLoram}. An alternative explanation involves the smearing of the critical temperature due to sample inhomogeneities or a density of states for superconducting electrons that is too small compared to the density of one-electron excitations associated with spin-density wave order \cite{Balatsky}.

\subsection{Specific-heat calculations for zero-dimensional superconductors}
\noindent

Generally, classical 0D superconductors, also referred to as classical or superconducting dots, account for information storage and processing as discrete bits in classical binary states (either 0 or 1) and operate according to classical physics \cite{Timm}. In contrast, quantum 0D superconductors, or quantum dots, are nanoscale superconductors in which electrons are confined, enabling quantum-mechanical behavior \cite{roger4,Lakic}. This behavior allows quantum dots to exist in superposition, representing both 0 and 1 simultaneously, and to exhibit phenomena such as entanglement \cite{Danga}. These properties indicate potential advancements in computation and information processing. In fluctuating 0D superconductors, the SHJ is smeared in finite-volume systems, and all superconducting effects arise solely from fluctuation-induced superconductivity. By setting $\phi_0 = \phi\sqrt{V}$ and applying a Hartree-Fock calculation derived from the coarse-graining of a microscopic model (effective Landau theory), the integral over $\phi^{2}_0$ and $\phi^{4}_0$ in the partition function can be expressed in terms of the error function \cite{roger4, Tsague}. Given the correlated and fluctuating nature of 0D superconductors, the fluctuation component of the heat capacity near the critical point can be expressed as follows.
\begin{equation}
\delta{C}(0D, T) \simeq \frac{1}{V}\Bigg(1 + T_{c_0}\frac{\partial{\upsilon(d, T)}}{\partial{T}}\Bigg)^2.
\end{equation}
\begin{figure}
 	\centerline{\includegraphics[width=8cm]{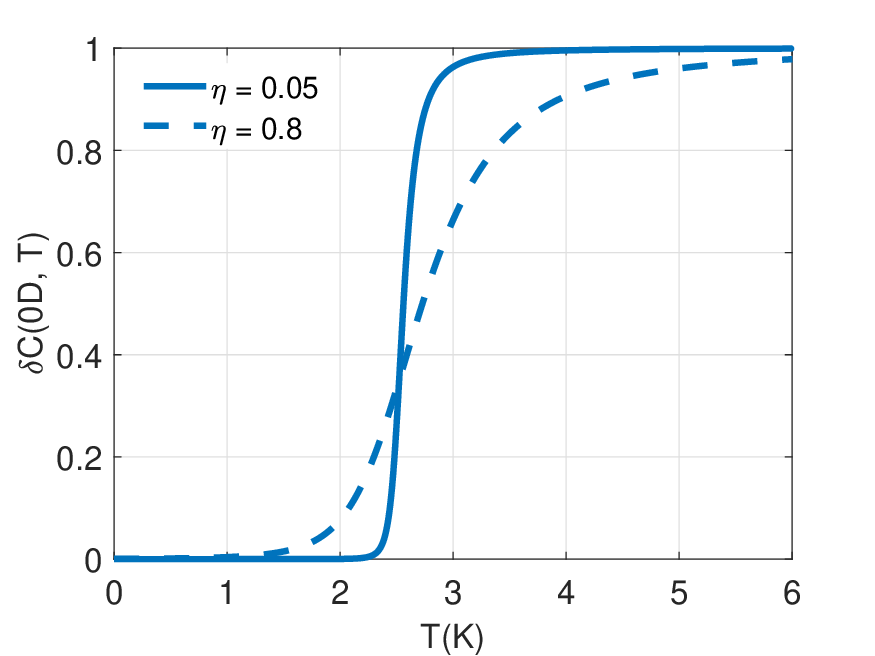}}
 	\caption{Plot of the zero-dimensional scaled specific heat per unit volume versus temperature according to Eq. (17). For $T_{c_0} = 2.5 K$, the fluctuation part of the specific heat $\delta{C}(0D, T) \rightarrow 0$ for $T \rightarrow 0$.}
 	\label{figure3}
 \end{figure}
Interestingly, Eq. (17), as depicted in figure 5, demonstrates that increased fluctuations in the superconducting OP reduce the lifetime of Cooper pairs. Analysis of the zero-field specific heat identifies two competing ground states: a conducting state with a large number of modes and a superconducting state with virtually no modes. When the 0D superconductor is cooled below the critical temperature $T_{c_0}$, electrons participating in pairing undergo charge transfer as single-electron excitations. The population of electrons involved in superconductivity decreases exponentially as the temperature approaches $0 K$, rather than exhibiting an abrupt decline at $T_{c_0}$ as observed in standard Landau theory. Single-electron excitations serve as direct evidence of spin-density wave order induced by thermal fluctuations \cite{Vavilov}. The gap region on both sides of $T_{c_0}$ is sharply defined in standard Landau theory ($\eta \rightarrow 0$) but becomes broader in the renormalized theory ($\eta > 0$). At both high and low temperatures, $\delta{C}(0D, T)$ is associated with quantum criticality and deviates from conventional behavior as $\eta$ increases. The existence of the gap region prevents the coexistence of superconducting and spin-density wave orders, thereby accounting for the absence of a SHJ in 0D systems. 

\section{conclusion}
\noindent

The influence of pairing mechanisms and one-electron excitations on the thermodynamic properties of high-$T_{c_0}$ superconductors is examined. A renormalized phenomenological approach is presented to analyze the research landscape, with particular focus on the role of dimensionality in characterizing superconductors under zero external fields and enhancing their properties. The specific heat jump (SHJ) is identified as a key indicator for distinguishing superconductors and determining the true critical temperature at which superconductivity emerges. This jump arises from the abrupt loss of normal electron excitations (quasiparticles), resulting in reduced thermal energy and a marked increase in specific heat. 
Notably, this increase does not occur in zero- and one-dimensional systems, which exhibit distinct shapes and characteristic properties due to finite energy levels. The existence of a gap region prevents the coexistence of spin-density wave and superconducting orders, explaining the absence of a specific heat jump in these low-dimensional systems. The interplay between thermal fluctuations and mass renormalization near the critical point significantly influences the power-law behavior of the SHJ. The proposed model offers a universal framework for analyzing rounding phenomena and the fluctuation-dependent linear, quasi-logarithmic, and quasi-exponential divergence of the specific heat jump. 

Adjusting renormalized quadratic coefficients to account for system dimensionality, while acknowledging that fluctuation effects intensify as dimensionality decreases, supports and reaffirms the Mermin-Wagner-Hohenberg theorem, particularly for $d \leq 2$. The dimension $d_c = 2$ is identified as a threshold below which order parameter fluctuations become sufficiently pronounced to dominate over thermal fluctuations. The evolution of SHJs with varying dimensionality results in the complete disappearance of the SHJ at $T_c$ in families of high-temperature superconductors such as BiSrCaCuO. In YBCO, synthesized as YBa$_2$Cu$_3$O$_{7 - \delta}$ with $0 \leq \delta \leq 0.18$, a transition occurs from relatively weak three-dimensional fluctuations at $\delta = 0$ to strong two-dimensional critical fluctuations over a broad temperature range for $\delta \geq 0.1$. 
The model demonstrates that the crossover from three-dimensional to two-dimensional behavior is accompanied by a reduction in the specific heat jump. An observed increase in this jump for low values of $\varepsilon$ is part of a complex process in which rate-domain lifetimes increase as the system transitions from a homogeneous to a heterogeneous regime. Rather than characterizing the specific heat jump with a power law, this study adjusts the internal model coefficients based on the scale of observation.

\section*{Acknowledgments}

\noindent
This work was supported by the Organization for Women in Science for the Developing World ({\color{blue}OWSD}) and the Swedish International Development Cooperation Agency ({\color{blue}SIDA}). 

\bigskip

{\color{blue} References}

\end{document}